\documentclass[a4paper,11pt]{article}
\pdfoutput=1 

\usepackage{jheppub} 

\usepackage[T1]{fontenc} 
\usepackage{comment}
\usepackage{color}
\usepackage{todonotes}
\usepackage{ulem}
\usepackage[thicklines]{cancel}

\newcommand{\tr}{\text{tr}}
\newcommand{\bra}{\langle}
\newcommand{\ket}{\rangle}
\newcommand{\eq}[1]{\begin{align}#1\end{align}}

\newcommand{\A}{{\cal A}}
\newcommand{\bO}{\boldsymbol{O}}
\newcommand{\bI}{\boldsymbol{I}}
\newcommand{\bJ}{\boldsymbol{J}}

\title{\boldmath Thermal Corrections to R\'enyi Entropy in BMS Field Theory}

\author[a]{Yuan Zhong}

\affiliation[a]{International Centre for Theoretical Physics Asia-Pacific (ICTP-AP),\\University of Chinese Academy of Sciences (UCAS), Beijing, China}

\emailAdd{zhongyuan@ucas.ac.cn}

\abstract{In the study of three-dimensional flat holography, the BMS field theory manifests the infinite-dimensional BMS$_3$ symmetry, a powerful tool in elucidating numerous universal phenomena. This paper explores a certain low-temperature limit of the BMS field theory. The primary focus lies in the calculation of the thermal correction to the R\'enyi entropy of the single interval on the cylinder from the replica trick and the uniformizing map. As a double check, an alternative method calculating the entanglement entropy is introduced, with the entanglement first law and the modular Hamiltonian.}

\begin{document} 
\maketitle
\flushbottom

\section{Introduction}\label{sec:intro}

In the pursuit of understanding quantum gravity, the holographic principle has emerged as a profound concept, establishing a deep-seated connection between $(d+1)$-dimensional quantum gravity and $d$-dimensional quantum field theory \cite{tHooft:1993dmi}. A pinnacle realization of this principle is the AdS/CFT correspondence \cite{Maldacena:1997re,Gubser:1998bc,Witten:1998qj}, which aligns quantum gravity in $(d+1)$-dimensional asymptotically anti-de Sitter (AdS) spacetime with a $d$-dimensional conformal field theory (CFT) on the asymptotic boundary. Central to this correspondence is the notion of symmetry. The asymptotic symmetry of the bulk theory equates the symmetry of the boundary theory, providing a powerful framework for deriving universal results and constraints.

In the study of holographic description of asymptotically flat gravity, inspired by the success of the role of asymptotic symmetry played in the AdS/CFT correspondence, the study of the asymptotic symmetry in the asymptotically flat spacetime, known as the Bondi–van der Burg–Metzner–Sachs symmetry \cite{Bondi:1962px,Sachs:1962wk}, receives much interest in the last few years. A simpler version of the asymptotically flat gravity is the three-dimensional BMS$_3$ symmetry. Based on the BMS$_3$ symmetry, the three-dimensional flat holography was proposed \cite{Bagchi:2010zz,Bagchi:2012cy} that the three-dimensional asymptotic flat gravity is holographically described by a two-dimensional quantum field theory governed by the BMS$_3$ symmetry, known as the BMS field theory (BMSFT) or Carrollian conformal field theory, since the BMS$_3$ algebra is isomorphic to the Carrollian conformal algebra. This is an infinite-dimensional algebra, and the constraints from it lead to powerful constraints in the study of the BMS field theories.

In the AdS/CFT framework, an important probe of the duality involves the holographic entanglement entropy. The Ryu-Takayanagi formula \cite{Ryu:2006bv,Ryu:2006ef} establishes a correspondence between the entanglement entropy in the boundary and the area of a minimal surface in the bulk. In the case of the flat holography, an analogous formula has been proposed \cite{Bagchi:2014iea,Basu:2015evh,Hosseini:2015uba,Jiang:2017ecm,Hijano:2017eii,Godet:2019wje,Fareghbal:2019czx,Apolo:2020bld}. On the BMS field theory side, the entanglement entropy for a single interval on the cylinder or plane in the vacuum state can be obtained through the replica trick \cite{Bagchi:2014iea}.

While the entanglement entropy serves as a good measurement of entanglement in pure states, in practice, however, it is always thermally polluted. This paper addresses the entanglement entropy for a single interval in the thermal state — a task that is challenging in general given the existence of both the thermal circle and the spatial circle. However, in the low-temperature limit $\beta_\phi \gg L,{\beta_u}/{\beta_\phi} \leq O(1)$, where $L$ represents the circumference of the cylinder coordinated by $\phi$ and $u$, and $\beta_\phi$ and $\beta_u$ denote the lengths of the thermal circle along the $\phi$- and $u$-directions respectively, the leading thermal correction to the Rényi entropy becomes computationally feasible. Inspired by universal results in the low-temperature limit in Conformal Field Theory (CFT) \cite{Cardy:2014jwa,Herzog:2014fra,Herzog:2014tfa,Herzog:2015cxa}, we employ the replica trick to express the leading contribution in the thermal correction as a correlation function on the branched covering space, subsequently resolving it with the help of the uniformizing map.

The resulting leading thermal correction to the Rényi entropy takes a universal form
\eq{
\delta S_n =\frac{n}{1-n}\left[\left( \frac{\sin\frac{\pi l_\phi}{L}}{n\sin\frac{\pi l_\phi}{n L}} \right)^{2\Delta} e^{\frac{2 \pi l_u \xi}{L} \left( \cot\frac{\pi l_\phi}{L} -\frac{1}{n}\cot\frac{\pi l_\phi}{nL} \right) } -1\right] e^{-\frac{2\pi\beta_\phi\Delta}{L} -\frac{2\pi\beta_u\xi}{L}},
}
which solely dependent on the scaling dimension $\Delta$ and the boost charge $\xi$ of the first excited state, together with the geometric configuration of the entanglement interval including the ranges $l_\phi$ and $l_u$ of the interval along the $\phi$- and $u$-directions. The thermal correction to the entanglement entropy is obtained through $\delta S_E = \delta S_{n\to 1}$.

As a double check, we also employ the entanglement first law, which translates the calculation of the variation $\delta S_E $ of the entanglement entropy into the variation $\delta \bra K_{\A}\ket$ of the expectation value of the modular Hamiltonian. The latter can be computed directly since the modular Hamiltonian for a single interval on the cylinder in the thermal state can be written explicitly. Our calculations demonstrate a consistent agreement between these two approaches.

This paper is organized as follows. In Sec.~2, we provide a quick review of BMS field theory. In Sec.~3, we calculate the thermal correction to the Rényi entropy in a certain low-temperature limit using the replica trick and the uniformizing map. Additionally, we present an alternative method to compute the thermal correction to the entanglement entropy through the modular Hamiltonian and the entanglement first law, serving as a double-check. We conclude in Sec.~4 with a summary and outline potential future directions.

\section{Review on the BMS Field Theory}\label{sec:reviw}
In this section, we give a quick review on some aspects of the BMS field theory.
\paragraph{$\bullet$ BMSFT on the cylinder}
~\\
A BMSFT on a cylinder $(\phi,u)$ with a circumference
\eq{
\phi \sim \phi+L}
is a two-dimensional quantum field theory that is invariant under the following BMS transformations
\eq{
&\phi \to f(\phi),\\
&u \to f'(\phi) u +g(\phi).
}
Here, $f(\phi)$ and $g(\phi)$ are periodic functions in $\phi$ with the periodicity $L$. Then, the infinitesimal BMS transformation generators are obtained by taking the Fourier modes
\eq{
l_n&= i \frac{L}{2\pi} e^{i n \frac{2\pi}{L}\phi}\partial_\phi -n e^{i n \frac{2\pi}{L}\phi} u\partial_u,\\
m_n&=i \frac{L}{2\pi} e^{i n \frac{2\pi}{L}\phi}\partial_u.
}

\paragraph{$\bullet$ BMSFT on the plane}
~\\
The BMSFT on the $(x,y)$-plane is obtained from the following plane-to-cylinder transformation \cite{Bagchi:2015nca}
\eq{
\label{map:plane2cyl}
x &=e^{\frac{2\pi i }{L} \phi},\\
y &= \frac{2\pi i }{L}  e^{\frac{2\pi i}{L} \phi} u.	\nonumber
}

The infinitesimal symmetry generators on the plane are
\eq{
l_n &=-x^{n+1}\partial_x -(n+1) x^n y \partial_y,\\
m_n&= -x^{n+1}\partial_y.
}
They form the BMS algebra without a central term via the Lie bracket
\eq{
[l_n ,l_m] &=(n-m) l_{m+n},\\
[l_n, m_m] &=(n-m) m_{m+n},\\
[m_n, m_m]&=0.
}
At the quantum level, these symmetry generators $l_n$ and $m_n$ will become operators $L_n$ and $M_n$ which act on the state space. They form the BMS algebra with central charges $c_M$ and $c_L$ as
\eq{
[L_n ,L_m] &=(n-m) L_{m+n} +\frac{c_L}{12}n(n^2-1)\delta_{m+n},\\
[L_n, M_m] &=(n-m) M_{m+n}+\frac{c_M}{12}n(n^2-1)\delta_{m+n},\\
[M_n, M_m]&=0.
}

A primary operator $\psi$ of the boost charge $\xi$ and the conformal dimension $\Delta$ is specified by the following conditions
\eq{
[L_0, \psi] &=\Delta \psi,\\
[M_0,\psi] &=\xi \psi,\\
[L_n, \psi] &=0,~ n>0,\\
[M_n, \psi] &=0,~ n>0.
}
Under a BMS transformation 
\eq{
\tilde{x} &=f(x), \label{bms_trans_finite}\\
\tilde{y} &= f'(x)y +g(x),\label{bms_trans_finite_y}
}
a primary operator $\psi$ transforms as
\eq{
\tilde\psi (\tilde{x},\tilde{y}) =(f')^{-\Delta} e^{-\xi \frac{y f'' +g'}{f'}}\psi(x,y).
}

On the plane, the currents $J(x)$ and $P(x)$ admit the following mode expansions
\eq{
J(x) &= \sum_n L_n x^{-n-2},\\
P(x) &=\sum_n M_n x^{-n-2}.
}
Under the BMS transformation \eqref{bms_trans_finite} and \eqref{bms_trans_finite_y}, the currents $J(x)$ and $P(x)$ transform as \cite{Jiang:2017ecm}
\eq{\label{current_trans}
\tilde{P}(\tilde{x}) &=\left( \frac{\partial f}{\partial x} \right)^{-2} \left( P(x) -\frac{c_M}{12}\{f,x\} \right),\\
\tilde{J}(\tilde{x}) &=\left( \frac{\partial f}{\partial x} \right)^{-2} \left( J(x) -\frac{c_L}{12}\{f,x\} \right) + \left( \frac{\partial g}{\partial x} \right)^{-2} \left( P(x) -\frac{c_M}{12}\{g,x\} \right) \nonumber.
}

\paragraph{$\bullet$ State-operator correspondence}
~\\
On the $(x,y)$-plane, the in-state corresponds to an operator inserted at $x=0$ \cite{Hao:2021urq}. From the plane-to-cylinder map \eqref{map:plane2cyl}, in the cylinder coordinate, the in-state is inserted at $\phi=i\infty$. Similarly, the out-state is inserted at $\phi=-i\infty$ in the cylinder coordinate.

\paragraph{$\bullet$ Vacuum state}
~\\
In particular, the above discussion can be specified to the vacuum state $|0\ket$ which is the highest weight state with vanishing conformal dimension $\Delta$ and boost charge $\xi$.
\eq{
L_n |0\rangle =0, ~ n \geq 0,\\
M_n |0\rangle =0, ~ n \geq 0.
}
In the following of the paper, we will assume that the vacuum state is unique. Also, to make sense of the first excited state, we will assume that the spectrum of the primaries is bounded from below. Moreover, for the first excited state to exist and dominate, we will constrain to theories satisfying the vacuum gap condition, which means the first excited state is gapped away from the vacuum state with finitely large conformal dimension and boost charge.

A prototype of counter examples that does not satisfy the above conditions could be the BMS free scalar with non-compact scalar field $\Phi$. In this case, the boost charge $\xi=-\frac{\alpha^2}{2}$ of the vertex operator $e^{i \alpha \Phi}$ could be either positive or negative with $\alpha \in \mathbb{R}\cup i\mathbb{R}$ hence the spectrum is not bounded from below. Also, with $\alpha$ small enough, the boost charge could be as close to the $0$ as one demonstrates, hence the vacuum state is not separable from other states as well.

However, when one consider a compact free scalar $\Phi \in S^1$, the above assumptions about the vacuum states are satisfied. Also, for a BMS field theory with finitely many primaries \cite{Yu:2022bcp,Hao:2022xhq}, the above assumptions are satisfied as well. At the cost of mild loss of the universality, in the following of the paper, we will constrain to BMS filed theories with a unique separable vacuum state and a spectrum that is bounded from below.

\section{Thermal Corrections to the Rényi  Entropy}
In this section, we use the replica trick and the uniformizing map to calculate the thermal correction to the R\'enyi entropy in the BMSFT for a single interval $\A$ on the cylinder with circumference $L$.

\subsection{Thermal Corrections to Rényi Entropy in CFT$_2$}
Before we continue our calculation of the thermal correction to the R\'enyi entropy for a single interval on the cylinder in the BMSFT, we would like to first review the similar calculation in the case of CFT$_2$ \cite{Cardy:2014jwa} first.

We assume that the theory is put on a cylinder with the circumference $L$, coordinatized by $w=x-it$, the thermal density matrix written in terms of a complete set of states is
\eq{
\rho =\frac{1}{\tr(e^{-\beta H})} \sum_{|\phi\ket} |\phi\ket \bra \phi| e^{-\beta E_{\phi}}.
}
The Hamiltonian on the cylinder in the CFT is the combination of the left- and the right-moving zeroth-level Virasoro generators and the central charge,
\eq{
H =\frac{2\pi}{L} \left( L_0 +\bar{L}_0 -\frac{c}{12} \right).
}
Here, we have assumed that $c_L=c_R=c$. With the assumptions that there exists a unique ground state $|0\ket$, and that the spectrum of conformal dimensions $\Delta=h+\bar{h}$ is positive and gapped from the smallest positive value, there should exist an operator $\psi$ of conformal weights $(h,\bar{h})$ carrying this smallest $\Delta$. This $\psi$ has the smallest energy $E_{\psi}=\frac{2\pi}{L}(\Delta -\frac{c}{12})$. Then, in the low-temperature limit $\beta\gg L$, the thermal density matrix admits the following expansion
\eq{\label{cft_rho}
\rho = \frac{|0\ket \bra 0| +|\psi\ket \bra\psi|e^{-2\pi\Delta\beta/L}+\cdots}{1 +e^{-2\pi\Delta\beta/L} +\cdots}.
}
We consider the entanglement region to be a single interval $\A$ with two endpoints 
\eq{
\partial_- \A: w=\bar{w}=w_1, ~ \partial_+ \A: w=\bar{w}=w_2.
}
For convenience, we also introduce the rescaled endpoints
\eq{
\theta_{1,2} = \frac{2\pi w_{1,2}}{L}
}
and their difference
\eq{
l=w_2-w_1.
}
Recall the Renyi entropy is defined by
\eq{\label{def:renyi}
S_n = \frac{1}{1-n} \log \tr(\rho_A^n).
}

So we need to calculate the trace of the $n$-th power of the reduced density matrix $\rho_\A$. This can be expanded according to the expansion \eqref{cft_rho} of the thermal density matrix as
\eq{
\tr\rho_A^n &= \frac{\tr \left[ \tr_{\bar{\A}}(|0\ket \bra 0| +|\psi\ket \bra\psi|e^{-2\pi\Delta\beta/L}+\cdots) \right]^n}{(1 +e^{-2\pi\Delta\beta/L} +\cdots)^n}\\
&=\tr(\tr_{\bar{\A}} |0\ket \bra 0|)^n \bigg[1+ \bigg( \frac{\tr (\tr_{\bar{\A}} |\psi\ket\bra \psi| (\tr_{\bar{\A}} |0\ket \bra 0|)^{n-1})}{\tr (\tr_{\bar{\A}} |0\ket \bra 0|)^n} -1 \bigg) n e^{-2\pi\Delta\beta/L} +\cdots\bigg].	\label{trrhoAcft}
}

The first term $\tr(\tr_{\bar{\A}} |0\ket \bra 0|)^n$ is just the zero-temperature R\'enyi entropy. And the expression in the second term
\eq{
\frac{\tr (\tr_{\bar{\A}} |\psi\ket\bra \psi| (\tr_{\bar{\A}} |0\ket \bra 0|)^{n-1})}{\tr (\tr_{\bar{\A}} |0\ket \bra 0|)^n},
}
which determines the leading thermal correction, can be recasted as a 2-point function of the operator $\psi(w)$ on an $n$-sheeted copy $C_n$ of the cylinder branched over $\A$ via the state operator correspondence $|\psi\ket \sim \lim_{t\to-\infty}\psi(x,t)|0\ket$ and $\bra\psi | \sim \lim_{t\to \infty}\bra 0|\psi(x,t)$ as 
\eq{\label{whatever}
\frac{\tr (\tr_{\A} |\psi\ket\bra \psi| (\tr_{\A} |0\ket \bra 0|)^{n-1})}{\tr (\tr_{\A} |0\ket \bra 0|)^n} =\lim_{t_2 \to \infty, t_1 \to -\infty}\frac{\bra\psi(w_2,\bar{w}_2)\psi(w_1,\bar{w_1}) \ket_{C_n}}{\bra\psi(w_2,\bar{w}_2)\psi(w_1,\bar{w_1}) \ket_{C_1}}.
}

To calculate the 2-point function $\bra\psi(w_2,\bar{w}_2)\psi(w_1,\bar{w_1}) \ket_{C_n}$ on the $n$-sheeted copy $C_n$, we can use the following uniformizing map
\eq{\label{uni_cft}
\zeta^{(n)} =\bigg( \frac{e^{2\pi i w/L} -e^{i\theta_2}}{e^{2\pi i w/L} -e^{i\theta_1}} \bigg)^{1/n}	
}
to send $C_n$ to the $\zeta$-plane. The 2-point function on a plane in the CFT is just
\eq{
\bra \psi(\zeta^{(n)}_2,\bar{\zeta}^{(n)}_2)\psi(\zeta^{(n)}_1,\bar{\zeta}^{(n)}_1) \ket =\frac{1}{(\zeta^{(n)}_{21})^{2h}(\bar{\zeta}^{(n)}_{21})^{2\bar{h}}}.
}
Mapping it back to the $n$-sheeted copy $C_n$ along the uniformizing map \eqref{uni_cft}, we obtain the expression of the 2-point function on $C_n$ as
\eq{
\bra\psi(w_2,\bar{w}_2)\psi(w_1,\bar{w_1}) \ket_{C_n} =\frac{(\frac{d \zeta_1}{d w_1}\frac{d \zeta_2}{d w_2})^h}{\zeta_{12}^{2h}} \frac{(\frac{d \bar{\zeta}_1}{d \bar{w}_1}\frac{d \bar\zeta_2}{d \bar{w}_2})^{\bar h}}{\bar\zeta_{12}^{2\bar{h}}}.
}
Substituting this into \eqref{whatever},we have
\eq{
\frac{\bra\psi(w_2,\bar{w}_2)\psi(w_1,\bar{w_1}) \ket_{C_n}}{\bra\psi(w_2,\bar{w}_2)\psi(w_1,\bar{w_1}) \ket_{C_1}} = \bigg[ \frac{1}{n^{2h}} \left( \frac{\zeta_1^{(n)} \zeta_2^{(n)}}{\zeta_1^{(1)}\zeta_2^{(1)}} \right)^{h} \left( \frac{\zeta_2^{(1)} -\zeta_1^{(1)}}{\zeta_2^{(n)} -\zeta_1^{(n)}} \right)^{2h} \Bigg] \cdot [\text{complex conjugate}].
}
After taking the limit $t_1\to -\infty$ and $t_2\to \infty$, we have
\eq{
\frac{\bra \psi(i\infty)\psi(-i\infty) \ket_{C_n}}{\bra \psi(i\infty)\psi(-i\infty) \ket_{C_1}} \frac{1}{n^{2\Delta}} =\left( \frac{\sin\frac{\theta_2 -\theta_1}{2}}{\sin\frac{\theta_2 -\theta_1}{2n}} \right)^{2\Delta}.
}
Then, from \eqref{trrhoAcft} and the definition of the R\'enyi entropy, we obtain the leading thermal correction to the R\'enyi entropy as
\eq{
\delta S_n = \frac{1}{1-n} \bigg( \frac{\sin^{2\Delta}(\pi l/L)}{n^{2\Delta-1}\sin^{2\Delta}(\pi l/nL)}-n \bigg) e^{-2\pi \Delta\beta/L}+ o(e^{-2\pi \Delta\beta/L}).
}
 
 In this calculation, suitable assumptions about the spectrum have been proposed so that the leading contribution to the thermal correction of the R\'enyi entropy is captured by the correlation function of the lightest operator on the branched covering space. The latter is further worked out with the help of the uniformizing map that sends this $n$-sheeted copy space to the plane.

\subsection{Thermal Correction Dominated by the Singlet Primary}\label{sec_singlet}
Consider a two-dimensional BMS filed theory on the cylinder coordinated by $(\phi, u)$ with circumference $L$
\eq{
\phi \sim \phi +L.
}

To introduce the temperature, we consider the following thermal identification \footnote{Here, we consider the case that $\beta_u$ takes the same sign as $\beta_\phi$, because we are going to assume the boost charge $\xi$ is bounded from below. If $\xi$ is bounded from above instead, then we should consider $(\phi, u) \sim (\phi +i\beta_\phi, u -i\beta_u )$ instead.}
\eq{
(\phi, u) \sim (\phi +i\beta_\phi, u +i\beta_u ).
}
Although it is most natural to consider a thermal circle along the $u$-direction along since the $u$-direction is the null time direction. However, as we will discuss in the Sect.~\ref{u-circle}, it is challenging to consider the low temperature expansion of the thermal state in this case since there are infinitely many $M_{-\vec{k}}$ descendants of the vacuum sharing the same boost charge with the vacuum hence they have the same effective energy conjugate to the $u$-direction. Instead, to fully exploit the power the BMS$_3$ symmetry, we will consider the thermal circle $(\beta_\phi,\beta_u)$ along both $\phi$- and $u$-directions for the low temperature limit.

The corresponding thermal density matrix is
\eq{\label{def:thermal_state}
\rho =\frac{ e^{-\beta_\phi L_0^{cyl} -\beta_u M_0^{cyl}} }{ \tr \left( e^{-\beta_\phi L_0^{cyl} -\beta_u M_0^{cyl}} \right) }.
}
Here, $L_0^{cyl}$ and $M_0^{cyl}$ are charges generating translations along $\phi$ and $u$ directions respectively. Under the plane-to-cylinder transformation of the currents \eqref{current_trans}, these cylinder translation generators are related to the canonical BMS generators $L_0$ and $M_0$ as
\eq{
L_0^{cyl} =\frac{2\pi}{L}( L_0-\frac{c_L}{24}), \quad M_0^{cyl}=\frac{2\pi}{L}( M_0-\frac{c_M}{24}).
}

Substituting this back into \eqref{def:thermal_state}, the thermal density matrix written in terms of  canonical BMS generators is
\eq{
\rho &=\frac{ e^{-\beta_\phi \frac{2\pi}{L}( L_0-\frac{c_L}{24}) -\beta_u \frac{2\pi}{L}( M_0-\frac{c_M}{24})} }{\tr\left( e^{-\beta_\phi \frac{2\pi}{L}( L_0-\frac{c_L}{24}) -\beta_u \frac{2\pi}{L}( M_0-\frac{c_M}{24})} \right)}\nonumber\\
&= \frac{ e^{-\beta_\phi \frac{2\pi}{L} L_0 -\beta_u \frac{2\pi}{L}M_0} }{\tr\left( e^{-\beta_\phi \frac{2\pi}{L}L_0 -\beta_u \frac{2\pi}{L}M_0} \right)}.
}

\paragraph{$\bullet$ Low Temperature Expansion}
~\\
We consider the BMSFT whose spectrum satisfies the following conditions so that the low-temperature expansion of the thermal density matrix is dominated by the first excited state.
\begin{itemize}
	\item[--] There exists a unique ground state $|0\rangle$, around which we can turn on a small temperature and expand the thermal density matrix.
	\item[--] In the spectrum both the conformal weight $\Delta$ and the boost charge $\xi$ are bounded from below. This is the analogue of the condition that the energy is bounded from below in usual Lorentzian field theories. With this assumption, the effective energy $-\beta_\phi \frac{2\pi}{L} L_0 -\beta_u \frac{2\pi}{L}M_0$ conjugate to the thermal circle is then bounded from below. As a result, it is possible to expand the thermal density matrix around the vacuum state.
	\item[--] There exists a gap between the ground state $|0\rangle$ and the lightest state $|\psi\rangle$ corresponding to the primary operator $\psi$ labelled by $(\Delta,\xi)$.
\end{itemize}
The last condition requires more explanation. As we turn on a small temperature, there might be several candidate lightest states above the ground state. Depending on the approach to the low-temperature limit, the operator $\psi$ with the smallest $\Delta +\frac{\beta_u}{\beta_\phi} \xi$ excites first.

There are still several difficulties to obtain an expansion dominated by $\psi$. First, due to the non-unitary nature, although $M_0$ is self-adjoint, it is not diagonalizable. For example, there are two descendants of $\psi$ at the level 1, $M_{-1}|\psi\rangle$ and $L_{-1}|\psi\rangle$. $M_0$ acts on them non-diagonally as a Jordan block
\eq{
M_0 \begin{pmatrix} M_{-1}|\psi\rangle \\ L_{-1}|\psi\rangle \end{pmatrix} = \begin{pmatrix} \xi & 0 \\ 1 & \xi \end{pmatrix}  \begin{pmatrix} M_{-1}|\psi\rangle \\ L_{-1}|\psi\rangle \end{pmatrix}.
}
As a consequence,  the thermal density matrix $\rho$ is also non-diagonalizable, and it is not possible to expand $\rho$ in terms of eigenstates $\{\Phi\}$ of $L_0$ and $M_0$ such as
\eq{
\rho \propto \sum_{\Phi} e^{-\frac{2\pi}{L} \left(\beta_\phi L_0^{\Phi} +\beta_u M_0^{\Phi} \right)} |\Phi\ket \bra\Phi| .
}
Another problem is that there are infinitely many descendants created by $M_{-k}$'s with the same boost charge $\xi$ as $\psi$ itself, because $M_{-k}$ all commute with $M_0$. So, in a low-temperature limit with $\beta_u \gg \beta_\phi$, these descendants will not be suppressed.

At this point, we will not try to answer the interesting question of the meaning of a non-diagonalizable density matrix. Instead, we restrict to a particular type of low-temperature limit to avoid the above difficulties.
\begin{itemize}
	\item[--] Consider the following low-temperature limit
\eq{\label{lowTlimit}
\beta_\phi \gg L,  \quad \frac{\beta_u}{\beta_\phi} \leq O(1).
} Then, the primary operator $\psi$ dominates the thermal density matrix expansion.
\end{itemize}

Under these assumptions, the thermal density matrix is dominated by $\psi$ at this low temperature as
\eq{\label{lowTexpan:totalstate}
	\rho&= \frac{|0\rangle \langle 0| +|\psi\rangle \langle \psi| e^{-\frac{2\pi\beta_\phi}{L}\Delta -\frac{2\pi\beta_u}{L}\xi} +\cdots}{1 +e^{-\frac{2\pi\beta_\phi}{L}\Delta -\frac{2\pi\beta_u}{L}\xi} +\cdots}.
}\\

\paragraph{$\bullet$ Entanglement measurements} 
~\\
Consider the entanglement region $\A$, the reduced density matrix on $\A$ is
\eq{
\rho_\A = \tr_{\bar \A} \rho.
}
We are interested in the following entanglement measurements: the Renyi entropy
\eq{\label{def:renyi}
S_n = \frac{1}{1-n} \log \tr(\rho_A^n)
}
and the entanglement entropy
\eq{
S_E= -\tr \rho_A \log \rho_A =S_{n\to 1}.
}
~\\

Concretely, we consider the entanglement region to be a single interval $\cal A$ specified by its endpoints
\eq{
\partial_- {\cal A} =(\phi_-,u_-), \quad \partial_+{\cal A} =(\phi_+,u_+).
}
For convenience, let us introduce the range of the interval $\A$ in the $\phi$- and the $u$-directions as
\eq{
l_\phi =\phi_+ -\phi_-, ~ l_u=u_+ -u_-.
}

Under the above low-temperature expansion \eqref{lowTexpan:totalstate}, $\tr\rho_\A^n$ can be expanded as
\eq{\label{lowTexpan:A}
\tr \rho_\A^n &=\frac{ \tr[\tr_{\bar \A}({|0\rangle \langle 0| +|\psi\rangle \langle \psi| e^{-2\pi\beta_\phi\Delta/L -2\pi\beta_u\xi/L} +\cdots})]^n }{(1 +e^{-2\pi\beta_\phi\Delta/L -2\pi\beta_u\xi/L} +\cdots)^n}\\
&=\tr(\tr_{\bar \A} |0\rangle \langle 0|)^n \bigg[1+ \bigg( \frac{\tr [\tr_{\bar \A} |\psi\rangle\langle \psi| (\tr_{\bar \A} |0\rangle \langle 0|)^{n-1}]}{\tr (\tr_{\bar \A} |0\rangle \langle 0|)^n} -1 \bigg) n e^{-\frac{2\pi\beta_\phi}{L}\Delta -\frac{2\pi\beta_u}{L}\xi} +\cdots\bigg].
}
The first term $\tr(\tr_{\bar \A} |0\rangle \langle 0|)^n$ corresponds to the ground-state R\'enyi entropy. The second term determines the leading contribution to the low-temperature thermal correction. To calculate this term, we use the replica trick and the state-operator correspondence to replace it by a 2-point function of $\psi$ on the $n$-sheeted copy $C_n$ of the original space branched over $\partial \A$. Use the state-operator correspondence, the in-state $|\psi\rangle$ corresponds to
\eq{
|\psi\rangle=\lim_{\phi\to i\infty}\psi(\phi,u)|0\ket,
}
and the out-state $\bra\psi|$ corresponds to
\eq{
\bra\psi|=\lim_{\phi\to -i\infty}\bra 0|\psi(\phi,u).
}
Together with the replica trick, the coefficient in the thermal correction term can be written as
\eq{
\frac{\tr [\tr_{\bar \A} |\psi\ket\bra \psi| (\tr_{\bar \A} |0\ket \bra 0|)^{n-1}]}{\tr (\tr_{\bar \A} |0\ket \bra 0|)^n} &=\lim_{\substack{\phi_1\to +i\infty \\ \phi_2\to -i\infty}}\frac{\tr [\tr_{\bar \A}\left( \psi(\phi_1,u_1)|0\ket \bra 0|\psi(\phi_2,u_2) \right) (\tr_{\bar\A} |0\ket \bra 0|)^{n-1}]}{\tr (\tr_{\bar\A} |0\ket \bra 0|)^n}\\
&= \lim_{\substack{\phi_1\to +i\infty \\ \phi_2\to -i\infty}}\frac{\bra \psi(\phi_2,u_2) \psi(\phi_1,u_1) \ket_{C_n} }{\bra \psi(\phi_2,u_2) \psi(\phi_1,u_1) \ket_{C_1} }.	\label{correction-as-2pt}
}

~\\
Now, we can use the uniformizing map to calculate this 2-point function of $\psi$ on $C_n$.
\paragraph{$\bullet$ Uniformizing Map}
~\\
To calculate the 2-point function on $C_n$, we use the following uniformizing map from $C_n$ to the plane,
\eq{
	x &=\left(\frac{e^{2\pi i \phi/L} -e^{2\pi i \phi_-/L}}{e^{2\pi i \phi/L}-e^{2\pi i \phi_+/L}}\right)^{1/n} =: f^{(n)}(\phi) \label{uniformmap}\\
	y&= \left( u -\frac{l_u}{2\sin\frac{\pi l_\phi}{L}}\sin\frac{\pi( 2\phi -\phi_- -\phi_+)}{L} \right)\frac{d}{d \phi} f^{(n)}(\phi).	\nonumber
}
This transformation can be decomposed into several steps. In the $x$-direction, the plane-to-cylinder map $z=e^{2\pi i\phi/L}$ maps the $S^{1}$-coordinate $\phi$ to the analytically continued complex $z$-plane. Then, on this complex plane, the $z$-coordinate of $\partial \A$ becomes $z_{\pm}=e^{2\pi i \phi_{\pm}/L}$. To introduce the $n$-sheeted copy of this analytically continued space branched over $z_{\pm}$, we apply an $SL(2,\mathbb C)$ transformation $w=\frac{z-z_-}{z-z_+}$ which sends $z_-$ to $0$ and $z_+$ to $\infty$, and take the $n$-th root of it. In the $y$-direction, the subtraction $\left( u -\frac{l_u}{2\sin\frac{\pi l_\phi}{L}}\sin\frac{\pi( 2\phi -\phi_- -\phi_+)}{L} \right)$ cancels the range $l_u$ of the interval $\cal A$ in $u$-direction.\\

The 2-point function of the primary operators on the plane is determined by the symmetry up to a normalization factor $N$ \cite{Bagchi:2009pe,Chen:2019hbj},
\eq{\label{plane2pt}
\bra \psi(x_1,y_1)\psi(x_2,y_2) \ket =N x_{12}^{-2\Delta}e^{-2\xi \frac{y_{12}}{x_{12}}}.
}
Mapped to the cylinder coordinate along \eqref{uniformmap}, the primary operator $\psi$ transforms as
\eq{
\psi(\phi, u) &=\left( \frac{d x}{d \phi}\right)^{\Delta}e^{-\xi \frac{{y}\frac{d^2\phi}{dx^2}}{\frac{d\phi}{dx}} -\xi \frac{d}{d\phi}\left( \frac{l_u}{2\sin\frac{\pi l_\phi}{L}}\sin\frac{\pi( 2\phi -\phi_- -\phi_+)}{L} \right)} \psi(x,y)	\nonumber\\
&=f^{(n)'\Delta}(\phi) e^{-\xi \left( u f^{(n)'}(\phi)\frac{d(f^{(n)'}(\phi)^{-1})}{d\phi} +\frac{\pi l_u}{L \sin\frac{\pi l_\phi}{L}}\cos\frac{\pi( 2\phi -\phi_- -\phi_+)}{L}  \right)} \psi(x,y).
}
Thus, the correlation function on $C_n$ becomes
\eq{
&\bra \psi(\phi_2,u_2) \psi(\phi_1,u_1) \ket_{C_n} \nonumber\\
=& N f^{(n)'\Delta}(\phi_2) e^{-\xi \left( u_2 f^{(n)'}(\phi)\frac{d(f^{(n)'}(\phi_2)^{-1})}{d\phi_2} +\frac{\pi l_u}{L \sin\frac{\pi l_\phi}{L}}\cos\frac{\pi( 2\phi_2 -\phi_- -\phi_+)}{L}  \right)} \nonumber\\
&\times f^{(n)'\Delta}(\phi_1) e^{-\xi \left( u f^{(n)'}(\phi_1)\frac{d(f^{(n)'}(\phi_1)^{-1})}{d\phi_1} +\frac{\pi l_u}{L \sin\frac{\pi l_\phi}{L}}\cos\frac{\pi( 2\phi_1 -\phi_- -\phi_+)}{L}  \right)} x_{12}^{-2\Delta}e^{-2\xi \frac{y_{12}}{x_{12}}}.
}

Substituting this into \eqref{correction-as-2pt} and take the limit, we obtain the the correction term
\eq{
&\frac{\tr [\tr_{\bar \A} |\psi\ket\bra \psi| (\tr_{\bar \A} |0\ket \bra 0|)^{n-1}]}{\tr (\tr_{\bar \A} |0\ket \bra 0|)^n} = \lim_{\substack{\phi_1\to +i\infty \\ \phi_2\to -i\infty}}\frac{\bra \psi(\phi_2,u_2) \psi(\phi_1,u_1) \ket_{C_n} }{\bra \psi(\phi_2,u_2) \psi(\phi_1,u_1) \ket_{C_1} }\\
=&\left( \frac{\sin\frac{\pi l_\phi}{L}}{n\sin\frac{\pi l_\phi}{n L}} \right)^{2\Delta} e^{\frac{2 \pi l_u \xi}{L} \left( \cot\frac{\pi l_\phi}{L} -\frac{1}{n}\cot\frac{\pi l_\phi}{nL} \right) }.
}
In the explicit calculation, to take the limit $\phi_1\to +i\infty, \phi_2\to -i\infty$, we have set $\phi_1=i T_1$ and $\phi_2 =-i T_2$ and expand the above in order of $\epsilon_1=e^{-2\pi T_1/L}$ and $\epsilon_2=e^{-2\pi T_2/L}$.

Substituting this back to the definition \eqref{def:renyi} of the R\'enyi entropy, we obtain the thermal correction to the R\'enyi entropy
\eq{
\delta S_n =\frac{n}{1-n}\left[\left( \frac{\sin\frac{\pi l_\phi}{L}}{n\sin\frac{\pi l_\phi}{n L}} \right)^{2\Delta} e^{\frac{2 \pi l_u \xi}{L} \left( \cot\frac{\pi l_\phi}{L} -\frac{1}{n}\cot\frac{\pi l_\phi}{nL} \right) } -1\right] e^{-\frac{2\pi\beta_\phi\Delta}{L} -\frac{2\pi\beta_u\xi}{L}}.
} 

The thermal correction to the entanglement entropy can be obtained by taking the $n\to 1$ limit,
\eq{\label{deltaSE}
\delta S_E =\left[ 2\Delta(1-\frac{\pi l_\phi}{L}\cot\frac{\pi l_\phi}{L}) + 2 \xi \left( \frac{\pi^2 l_u l_\phi}{L^2 \sin^2\frac{\pi l_\phi}{L}}-\frac{\pi l_u}{L}\cot\frac{\pi l_\phi}{L} \right) \right]  e^{-\frac{2\pi\beta_\phi\Delta}{L} -\frac{2\pi\beta_u\xi}{L}}.
}

For a pure state, $S_n(\A)=S_n(\bar\A)$. However, the thermal correction contribution violates this equality. The compliment of $\A$ is an interval of range $L-l_\phi$ in the $\phi$-direction and $-l_u$ in the $u$-direction. Since $\delta S_n(L-l_\phi,-l_u)\neq \delta S_n (l_\phi,l_u)$, the R\'enyi entropy is indeed thermally polluted.

\subsection{Thermal Correction Dominated by the Multiplet Primary}
Previously, we obtained the thermal correction to the R\'enyi entropy and the entanglement entropy in the case that a singlet primary dominates the thermal correction. Now, we consider the case that a multiplet primary dominates the thermal correction. As we will see, the thermal correction to the R\'enyi entropy is just that of a singlet multiplied by the rank of the multiplet. However, this seemingly intuitive result is not that trivial. Actually, the off-diagonal terms dominate the expansion of the thermal density matrix, but they just do not contribute to the thermal correction to the R\'enyi entropy.

The $M_0$ acts on a rank-$r$ primary multiplet $\bO=(O_0,O_1,\cdots,O_{r-1})^{\text{T}}$ as \cite{Bagchi:2009ca,Chen:2020vvn,Chen:2022jhx}
\eq{
M_0 | O_a \ket &= \xi | O_a \ket + | O_{a-1} \ket, ~a=1,\cdots,r-1,\\
M_0 | O_0 \ket &= \xi | O_0 \ket, ~ a=0.
}
Or in a more compact form, $M_0 \bO =(\xi \bI_r  +\bJ_r) \bO$. Here, $\bI_r$ is the rank-$r$ identity matrix, and $\bJ_r$ is the rank-$r$ Jordan cell
\begin{equation}\label{jordan}
\bJ_r=
\begin{pmatrix}
 0& & & \\
  1& 0& & \\
  & \ddots&\ddots &\\
   & & 1& 0\\
\end{pmatrix}_{r\times r},
\end{equation}
which is nilpotent $(\bJ_r)^r=0$. The action of $e^{-\beta_\phi \frac{2\pi}{L}L_0-\beta_u \frac{2\pi}{L}M_0}$ on the primary part of this multiplet becomes $e^{-\beta_u \frac{2\pi}{L}\bJ_r} e^{-\frac{2\pi\beta_\phi}{L}\Delta -\frac{2\pi\beta_u}{L}\xi}$. The matrix part $e^{-\beta_u \frac{2\pi}{L}\bJ_r}$ can be expanded into finitely many terms as
\eq{\label{offdiag-rho}
e^{-\beta_u \frac{2\pi}{L}\bJ_r} =\sum_{k=0}^{r-1} \frac{(-\beta_u \frac{2\pi}{L})^k}{k!} (\bJ_r)^k.
}
Since $\beta_u \gg L$, it seems that the $k=r-1$ term dominates the expansion \eqref{offdiag-rho}. However, as we will see later, although this $\bJ^{r-1}$ term dominates the expansion of the matrix, it does not contribute to the thermal correction term to the R\'enyi entropy after taking trace. It is the $\bJ^{0}$ term that dominates the thermal correction. Explicitly, the $k=r-1$ term is
\eq{
\frac{(-\beta_u \frac{2\pi}{L})^{r-1}}{(r-1)!} (\bJ_r)^k= \frac{(-\beta_u \frac{2\pi}{L})^{r-1}}{(r-1)!} \begin{pmatrix}
0	&	&		&0\\
\vdots&\ddots				\\
0	&	&\ddots		&\\
1	&0	&\cdots	&0
\end{pmatrix}_{r\times r}
= \frac{(-\beta_u \frac{2\pi}{L})^{r-1}}{(r-1)!} |O_0\ket  \bra O_{r-1}^{\vee}|.
}
Here, the dual basis $\bra O_{a}^{\vee}|$ is defined by
\eq{
\bra O_{a}^{\vee} |O_b\ket =\delta_{a,b}.
}
Putting everything together, the multiplet version of the low-temperature expansion of density matrix \eqref{lowTexpan:totalstate} is
\eq{
	\rho&= \frac{|0\rangle \langle 0| +|O_0\ket \bra O_{r-1}^{\vee} | \frac{(-\frac{2\pi\beta_u}{L})^{r-1}}{(r-1)!} e^{-\frac{2\pi\beta_\phi}{L}\Delta -\frac{2\pi\beta_u}{L}\xi} +\cdots}{1 +r e^{-\frac{2\pi\beta_\phi}{L}\Delta -\frac{2\pi\beta_u}{L}\xi} +\cdots
}.
}
We can use the inner product between the in-state and the out-state within a multiplet \cite{Chen:2020vvn}
\eq{
\bra O_a | O_b \ket =\delta_{a+b,r-1}
}
to transfrom from the dual basis to the out-states,
\eq{
\bra O_{a}^{\vee} |=\bra O_{r-1-a}|.
}
Then, the density matrix can be written as 
\eq{
	\rho&= \frac{|0\rangle \langle 0| +|O_0\ket \bra O_{0}| \frac{(-\frac{2\pi\beta_u}{L})^{r-1}}{(r-1)!} e^{-\frac{2\pi\beta_\phi}{L}\Delta -\frac{2\pi\beta_u}{L}\xi} +\cdots}{1 +r e^{-\frac{2\pi\beta_\phi}{L}\Delta -\frac{2\pi\beta_u}{L}\xi} +\cdots}.
}

The correlation function \cite{Henkel:2013fja,Chen:2020vvn} among the rank-$r$ multiplet is
\begin{equation}\label{2pt}
\langle O_{a}(x_x,y_x)O_{b}(x_1,y_1)\rangle =\left\{\begin{array}{ll}
0& \mbox{for $q<0$}\\
\, \,  d_r\,  x_{12}^{-2\Delta_i} e^{-2\xi_i\frac{y_{12}}{x_{12}}}\frac{1}{q!}\left(-\frac{2y_{12}}{x_{12}}\right)^q,& \mbox{otherwise} \end{array} \right. ,~q=a+b-r+1.	
\end{equation}
In particular, for $r>1$
\eq{
\bra O_0(x,y) O_0(x',y') \ket =0.
}
So, we see this $\bJ^{r-1}$ term does not contribute to the thermal correction term at all. Moreover, it turns out that all the off-diagonal terms do not contribute to the leading correction to the R\'enyi entropy. To see this, consider the $\bJ^k$ summand in \eqref{offdiag-rho} written in the basis
\eq{
\bJ_r^k = \sum_{a=0}^{r-1-k} |O_a\ket \bra O_{a+k}^{\vee}| =\sum_{a=0}^{r-1-k} |O_a\ket \bra O_{r-1-a-k}|.
}
Since $q=(a) +(r-1-a-k) =r-1-k \geq r-1$ and the equality holds only if $k=0$,
the correlation function $\bra O_{a}(x,y) O_{r-1-a-k}(x',y')\ket$ vanishes for any $k>0$ because of \eqref{2pt}. Only for $k=0$, the correlation function does not vanish, i.e.,
\eq{
\bra O_a(x_2,y_2) O_{r-1-a}(x_1,y_1) \ket =N x_{12}^{-2\Delta} e^{-\xi\frac{y_{12}}{x_{12}}}, ~a=0,\cdots,r-1,
}
which is the same as the correlation function of a singlet \eqref{plane2pt}. So, the thermal correction to the R\'enyi entropy is just that of a singlet multiplied by $r$,
\eq{
\delta S_n =r\frac{n}{1-n}\left[\left( \frac{\sin\frac{\pi l_\phi}{L}}{n\sin\frac{\pi l_\phi}{n L}} \right)^{2\Delta} e^{\frac{2 \pi l_u \xi}{L} \left( \cot\frac{\pi l_\phi}{L} -\frac{1}{n}\cot\frac{\pi l_\phi}{nL} \right) } -1\right] e^{-\frac{2\pi\beta_\phi\Delta}{L} -\frac{2\pi\beta_u\xi}{L}}.
} 

We see that when a multiplet primary dominates the low-temperature expansion, although the off-diagonal contributions dominate the correction to the thermal density matrix, they do not contribute to the correction of the R\'enyi entropy. The result is just that of the singlet multiplied by the rank $r$. It will be interesting to find if there exist any other entanglement measurements to which the off-diagonal contributions do not vanish. We leave this to future work.

\subsection{Comments on Another Limit}\label{another limit}\label{u-circle}
So far, we consider the particular low-temperature limit \eqref{lowTlimit}, but there is also a complimentary choice to reach the low-temperature limit so that the boost charge $\xi$ dominates the first excited state. An extreme case is that the thermal circle is purely along the $u$-direction. The thermal circle is
\eq{
u \sim u+ i\beta_u, ~ \beta_u \gg L.
}
Then, the density matrix is proportional to $e^{-\beta_u M_0}$. In this case, any primary $\psi$ with boost charge $\xi>0$ is heavier than not only the vacuum state, but all the descendants of the vacuum (e.g., $M_{-\vec{k}}|0\ket$), because these descendants of the vacuum all have the boost charge $\xi=0$.

If in the spectrum the boost charge is gapped, then in the $\beta_u \gg L$ limit, the density matrix is dominated by the vacuum block, and all the vacuum descendants are just as heavy as the vacuum, thus a low-temperature expansion is hardly accessible. However, the result of such thermal correction to the entanglement entropy might be even more universal than the previous case, as it depends only on the vacuum block and the algebraic structure, not on the details of the spectrum.

On the other hand, if there exist any other primary operators with boost charge $0$, then the density matrix is dominated by these blocks together with the vacuum block. Since the operator $e^{-\beta_u M_0}$ does not care about the conformal weight at all, the results of the thermal correction might be similar to the case that only the vacuum block dominates.

To summarize, in this type of low-temperature limit, since all descendants of the vacuum are equally heavy measured by their boost charge, an honest calculation must include them all. Even it is still possible to expand the density matrix  $e^{-\beta_u M_0}$ organized by the orders of the Taylor expansion and the levels of the descendants, it is still hard to trace out $\bar{\A}$ and obtain the reduced density matrix on $\A$ in a workable way. However, since we expect the result to be universal, once we work it out in one explicit example, hopefully we might find a solution according to the answer. Currently, since this type of thermal circle is in BMSFT is still not well understood, we leave this to future work.

\subsection{Modular Hamiltonian Approach}

In this subsection, we calculate the thermal correction to the entanglement entropy from the modular Hamiltonian. As a double check, the result agrees with the previous calculation \eqref{deltaSE}. The modular Hamiltonian for the reduced density matrix on $\A$ is defined to be
\eq{
K_{\A} = -\log \rho_{\A}.
} From the entanglement first law, for an infinitesimal variation of the state, the calculation of the variation of the entanglement entropy can be replaced by the variation of the expectation value of the modular Hamiltonian
\eq{
\delta S_{\A}=\delta \bra K_{\A} \ket.
}
In general, the modular Hamiltonian $K_{\A}$ cannot be written down in terms of local data. Only in theories with enough symmetry the modular Hamiltonian has an explicit formula for simple entanglement regions and special states. In particular, the modular Hamiltonian in BMSFT \cite{Apolo:2020qjm} can be written down explicitly for a single interval on the cylinder under the vacuum state.

For the single interval $\A$ in the vacuum state on the cylinder with circumference $L$, the modular Hamiltonian $K_\A$ can be written as a local integral of the modular generator $\zeta_{\A}$ against the currents $J(\phi)$ and $P(\phi)$ as
\eq{\label{BMSmodH}
K_{\A} =\int_{\phi_-}^{\phi_+} d\phi \left[ \frac{L}{2\pi} \frac{\cos\frac{\pi l_\phi}{L}-\cos\frac{\pi(2\phi -\phi_+ -\phi_-)}{L}}{\sin\frac{\pi l_\phi}{L}} J(\phi) + \frac{l_u}{2 } \frac{\cot\frac{\pi l_\phi}{L}\cos\frac{\pi(2\phi -\phi_+ -\phi_-)}{L} -\csc\frac{\pi l_\phi}{L}}{\sin\frac{\pi l_\phi}{L}} P(\phi) \right].
}

To calculate the variation of the modular Hamiltonian, we need to calculate the variation of the currents $J(\phi)$ and $P(\phi)$,
\eq{
\delta\bra J \ket = \bra J \ket_{\rho} -\bra J \ket_{|0\ket},\\
\delta\bra P \ket = \bra J \ket_{\rho} -\bra P \ket_{|0\ket}.
}
Substitute the low-temperature expansion \eqref{lowTexpan:totalstate} of the thermal density matrix $\rho$, 
\eq{\label{var_currents}
\delta\bra J(\phi) \ket =e^{-\frac{2\pi\beta_\phi}{L}\Delta -\frac{2\pi\beta_u}{L}\xi} \left( \bra J(\phi) \ket_{|\psi\ket} - \bra J(\phi) \ket_{| 0\ket} \right),\\
\delta\bra P(\phi) \ket =e^{-\frac{2\pi\beta_\phi}{L}\Delta -\frac{2\pi\beta_u}{L}\xi} \left( \bra P(\phi) \ket_{|\psi\ket} - \bra P(\phi) \ket_{| 0\ket} \right). \nonumber
}

So, we need to calculate the difference of the expectation values of the currents between the primary state $|\psi\ket$ and the vacuum $| 0\ket$. For this, we apply the plane-to-cylinder transformation \eqref{map:plane2cyl} and insert the primary operator $\psi$ at the origin of the $(x,y)$-plane. Recall the mode expansion of the currents on the plane
\eq{
J(x) =\sum_n L_n x^{-n-2}, ~ P(x)=\sum_n M_n x^{-n-2}.
}
Thus, the expectation values of the currents on the plane under a primary state are
\eq{
\bra J^{pl}(x)\ket=x^{-2} \Delta, ~ \bra P^{pl}(x)\ket=x^{-2}\xi.
}
Applying the transformation of currents \eqref{current_trans}, the expectation values of currents on cylinder become
\eq{
\bra J(\phi) \ket &= \left( \frac{\partial x}{\partial \phi} \right)^2 J^{pl}(x) +\frac{c_L}{12}\{x,\phi\}=-\frac{4\pi^2}{L^2}\Delta +\frac{\pi^2}{L^2} \frac{c_L}{6},\\
\bra P(\phi) \ket&= \left( \frac{\partial x}{\partial \phi} \right)^2 P^{pl}(x) +\frac{c_M}{12}\{x,\phi\}=-\frac{4\pi^2}{L^2}\xi +\frac{\pi^2}{L^2} \frac{c_M}{6}.
}
Thus, the difference of the expectation values of the currents between $|\psi\ket$ and $|0\ket$ are
\eq{
\bra J(\phi) \ket_{|\psi\ket} - \bra J(\phi) \ket_{| 0\ket} &=-\frac{4\pi^2}{L^2}\Delta,\\
\bra P(\phi) \ket_{|\psi\ket} -\bra P(\phi) \ket_{|0\ket} &=-\frac{4\pi^2}{L^2}\xi.
}
Substituting this into \eqref{var_currents}, we obtain the variation of the currents
\eq{
\delta\bra J(\phi) \ket &=e^{-\frac{2\pi\beta_\phi}{L}\Delta -\frac{2\pi\beta_u}{L}\xi} \left( \bra J(\phi) \ket_{|\psi\ket} - \bra J(\phi) \ket_{| 0\ket} \right) =-\frac{4\pi^2}{L^2}\Delta \, e^{-\frac{2\pi\beta_\phi}{L}\Delta -\frac{2\pi\beta_u}{L}\xi} ,\\
\delta\bra P(\phi) \ket &=e^{-\frac{2\pi\beta_\phi}{L}\Delta -\frac{2\pi\beta_u}{L}\xi} \left( \bra P(\phi) \ket_{|\psi\ket} - \bra P(\phi) \ket_{| 0\ket} \right) =-\frac{4\pi^2}{L^2}\xi \, e^{-\frac{2\pi\beta_\phi}{L}\Delta -\frac{2\pi\beta_u}{L}\xi}.
}
For the modular Hamiltonian \eqref{BMSmodH}, the variation of the modular Hamiltonian is
\eq{
\delta \bra K_{\A} \ket
=&\int_{\phi_-}^{\phi_+} d\phi \left[ \frac{L}{2\pi} \frac{\cos\frac{\pi l_\phi}{L}-\cos\frac{\pi(2\phi -\phi_+ -\phi_-)}{L}}{\sin\frac{\pi l_\phi}{L}} \delta\bra J(\phi) \ket  + \frac{l_u}{2 } \frac{\cot\frac{\pi l_\phi}{L}\cos\frac{\pi(2\phi -\phi_+ -\phi_-)}{L} -\csc\frac{\pi l_\phi}{L}}{\sin\frac{\pi l_\phi}{L}} \delta\bra P(\phi) \ket  \right] \nonumber\\
=&\left[ 2\Delta(1-\frac{\pi l_\phi}{L}\cot\frac{\pi l_\phi}{L}) + 2 \xi \left( \frac{\pi^2 l_u l_\phi}{L^2 \sin^2\frac{\pi l_\phi}{L}}-\frac{\pi l_u}{L}\cot\frac{\pi l_\phi}{L} \right) \right]e^{-\frac{2\pi\beta_\phi}{L}\Delta -\frac{2\pi\beta_u}{L}\xi} .
}

This result agrees with the previous calculation \eqref{deltaSE} of the variation of the entanglement entropy.

\section{Discussion}

In this paper, we consider the single interval entanglement region on the cylinder in the BMSFT. We find a suitable low-temperature limit under which an expansion of the thermal density matrix dominated by the first excited operator is possible. In this limit, we calculate the thermal correction to the R\'enyi entropy by the replica trick and the uniformizing map. As a double check, for the thermal correction to the entanglement entropy, we also provide an alternative calculation by the modular Hamiltonian and the entanglement first law.

Though we provide a double check from another calculation of the entanglement entropy by modular Hamiltonian, it will be more satisfactory to have a numerical check in the concrete model. Despite the fact that several concrete BMSFT models have been found and studied recently, it seems that we still do not have a satisfactory understanding of their underlying Hilbert space structure and the correct way to discretize the models in a meaningful way. We leave this to future work until we have a better understanding of these concrete models. Also, a concrete model analysis might be helpful to understand another type of low-temperature limit in Sec.~\ref{another limit}.

Another interesting thing is to test this thermal correction term in the holographic entanglement proposals. For the finite temperature, the calculation on the cylinder is secretly on a torus, and the replica trick fails as the covering space is of high genus. However, a holographic calculation with temperature in the bulk is still possible using the geometric picture. Hence, a comparison between the low-temperature result in the bulk and boundary is possible.




\acknowledgments
I would like to thank Peng-xiang Hao, Wenxin Lai and Jun Nian for useful discussions. I would like to specially thank Jun Nian for proofreading the manuscript. This work was supported in part by the NSFC under grant No. 12147103.


\bibliographystyle{JHEP}
\bibliography{Note}



\end{document}